\documentclass[12pt]{book}

\usepackage[dvips]{graphicx,color}
\usepackage{makeidx,tsukuba}

\makeauthorindex
\makeindex

\begin{document}

\BookTitle{\itshape The 28th International Cosmic Ray Conference}
\CopyRight{\copyright 2003 by Universal Academy Press, Inc.}
\pagenumbering{arabic}

\chapter{
Search for a WIMP annihilation signature in the core of the globular cluster
M15}

\author{%
%
%
%
S.~LeBohec,$^{1,2}$ E.A.~Baltz,$^3$ I.H.~Bond, P.J.~Boyle, S.M.~Bradbury, J.H.~Buckley,
D.~Carter-Lewis, O.~Celik, W.~Cui, M.~Daniel, M.~D'Vali,
I.de~la~Calle~Perez, C.~Duke, A.~Falcone, D.J.~Fegan, S.J.~Fegan,
J.P.~Finley, L.F.~Fortson, J.~Gaidos, S.~Gammell, K.~Gibbs,
G.H.~Gillanders, J.~Grube, J.~Hall, T.A.~Hall, D.~Hanna, A.M.~Hillas,
J.~Holder, D.~Horan, A.~Jarvis, M.~Jordan, G.E.~Kenny, M.~Kertzman,
D.~Kieda, J.~Kildea, J.~Knapp, K.~Kosack, H.~Krawczynski, F.~Krennrich,
M.J.~Lang, E.~Linton, J.~Lloyd-Evans, A.~Milovanovic,
P.~Moriarty, D.~Muller, T.~Nagai, S.~Nolan, R.A.~Ong, R.~Pallassini,
D.~Petry, B.~Power-Mooney, J.~Quinn, M.~Quinn, K.~Ragan, P.~Rebillot,
P.T.~Reynolds, H.J.~Rose, M.~Schroedter, G.~Sembroski, S.P.~Swordy,
A.~Syson, V.V.~Vassiliev, S.P.~Wakely, G.~Walker, T.C.~Weekes,
J.~Zweerink \\
{\it
(1) Department of Physics and Astronomy, ISU, Ames, IA, 50011-3160, USA \\
(2) The VERITAS Collaboration--see S.P.Wakely's paper} ``The VERITAS
Prototype'' {\it from these proceedings for affiliations\\
(3) ISCAP, Columbia Astrophysics Laboratory, New York, NY 10027, USA }
}

\section*{Abstract}
The Whipple 10m Very High Energy gamma-ray telescope has been used to 
search
for indications of WIMP annihilation in the direction of the globular cluster
M15.  The upper limits derived constrain the amount of super-symmetric dark
matter that may reside in globular clusters.

\section{Introduction}
Studies of the CMB have accurately determined the density of dark matter
(non-baryonic clustering material) in the universe to be $\sim 23\%$ of the 
total
energy density, with most of the remainder being ``dark energy'' [14].  Weakly interacting massive particles (WIMPs) are an excellent dark
matter candidate, especially those arising in super-symmetric extensions to the
Standard Model.  Such particles can annihilate, and the gamma rays produced
should thus trace the dark matter density.  The best targets for the search of
annihilating dark matter are the most massive dense nearby structures. The
Galactic Center is an interesting candidate [16] and is being
observed above $\rm \sim 1~TeV$ with the Whipple telescope [2].
Other galaxies may be of interest as well but their large distances is a strong
penalty.  Relatively nearby dwarf galaxies [15] have been considered
as well as giant galaxies like M87 [1].  With mass to light
ratios from 0.5 to 2.5 [12], globular clusters are not a usual
target in searches for dark matter.  Their mass corresponding to the Jeans
scales at the time of recombination 
suggests that
they may have been the first structures to form and at least some of them
[13] may be relics of the galaxy formation epoch 
.  The cold dark matter driven structure formation scenario results in an
extended dark matter halo.  Dynamical studies have shown that if a dark matter
halo is attached to globular clusters, it does not extend beyond the stellar
distribution [11]. The halo is probably stripped by tidal effects when
the cluster passes through the disk and close to the bulge [6].  
Nevertheless some fraction of the original dark matter may remain
inside the cluster.  The density in the center of globular clusters typically
reaches $10^4M_\odot\;{\rm pc}^{-3}$ or even $10^5M_\odot\;{\rm pc}^{-3}$ in 
core collapsed clusters such as M15, with a strong radial density gradient.
If the dark matter traces the stellar distribution, it may yield interesting
annihilation rates even for small amounts of dark matter in the form of WIMPs.
  A different situation could arise if a few thousand solar
mass black hole occupied the center of the cluster as suggested by Gerssen et
al. [5].  The gamma ray emission would then depend on the history of the black 
hole formation and on the central dark matter evolution rate (for references 
see [17]).

For this preliminary analysis, we assumed the dark matter follows the observed 
stellar distribution to derive an expression for the expected gamma-ray flux 
from WIMP annihilation 
[9]. We describe the density profile as 
$\rho(r)={{\rho_0} \over {1+(r/R_c)^\alpha}}$ with a truncation at the tidal 
radius and $R_c$ being the core 
radius, $\alpha=2.2$ [18,19] and $\rho_0$ is normialized to reproduce the 
cluster 
mass within the tidal radius truncation. This is 
quite a conservative description as it underestimates the central density.
We have computed a gamma-ray flux  for the $119$ globular clusters tabulated 
by Gnedin et al. [6].  Restricting ourselves to declinations larger 
than $+10^\circ$, we find the most promising object to be NGC7078 (or M15)

\section{Very High Energy observations}

\begin{figure}[t]
  \begin{center}
    \includegraphics[height=15.5pc]{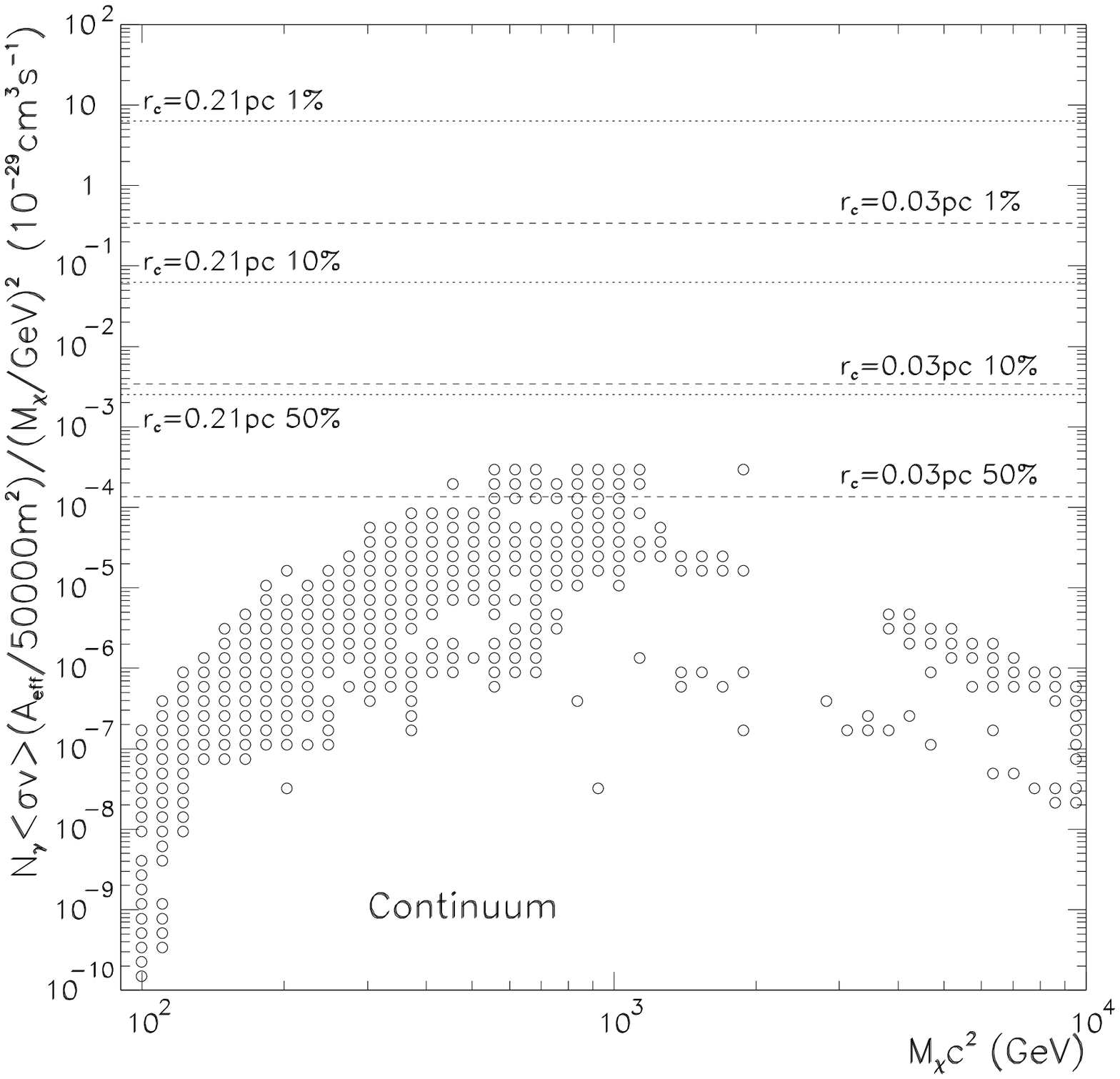}
    \includegraphics[height=15.5pc]{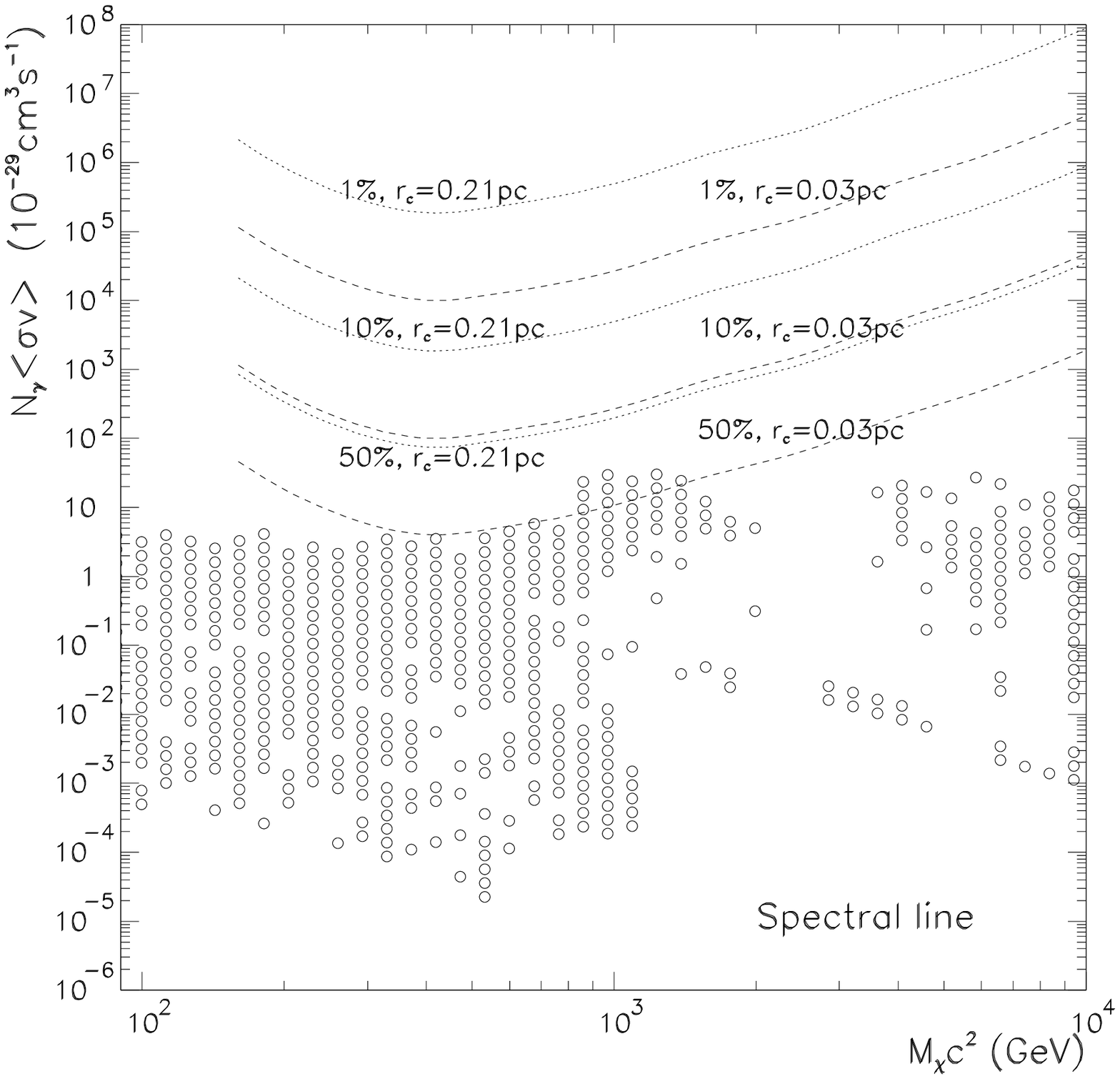}
  \end{center}
  \vspace{-0.5pc}
\caption{Estimated gamma-ray emission rate from WIMP annihilation and 
Whipple limits from observations of M15, as a function of the
WIMP mass.  A large number of super-symmetric models resulting from Monte-Carlo
 simulations as discussed in the text, are binned and then plotted for the 
continuum (left) and for the spectral line (right). For the continuum 
component, the annihilation rate is combined with the detector effective 
collection area ($A_{eff}$). For both plots, the curves represent the 99\%CL
upper limit we derived 
from our observation of M15 under different assumptions of the relative 
amount of super-symmetric dark matter (1\%, 10\% \& 50\%) and for different 
values of the core radius ($\rm 0.21pc$ \& $\rm0.03pc$). }
\end{figure}
In 2002, the Whipple 10m telescope was used to search for gamma-ray emission
from the globular cluster M15. The source was observed for 196 minutes. 
No significant excess was found and we derived the $99\%$ confidence level 
upper limit on the rate  to be $0.19\;{\rm min}^{-1}$.

The flux upper limit depends on the energy distribution of the putative
gamma-ray signal.  WIMP annihilation should yield two different spectral
components.  Spectral lines should result from processes $\chi \chi \rightarrow
\gamma \gamma$ and $\chi \chi \rightarrow Z_0 \gamma$ while a continuum is
produced by all the processes of type $\chi \chi \rightarrow X_1 X_2$ where
$X_1$ and $X_2$ are both particles which may decay or hadronize producing
secondary gamma rays.  A set of super-symmetric models was explored with the
DarkSUSY code [2,4,7], providing averaged gamma-ray emission rates from annihilation
$N_\gamma\langle\sigma v\rangle$ where $N_\gamma$ is the number of
gamma rays produced per annihilation and $\langle\sigma v\rangle$ is the
thermally averaged annihilation rate.  In the case of the continuum, the energy
of the gamma rays is different from the mass of the annihilating particles
and the effective collection area $A_{eff}$ of the detector estimated from Monte-Carlo
simulations, has to be folded in as it depends on the energy. The gamma-ray
emission rates are shown in figure 1 as a function of the WIMP mass $m_\chi$
for the continuum (left) and the spectral line (right).  
We converted our upper limit into
an upper limit of the gamma-ray production rate $N_\gamma\langle\sigma
v\rangle$ to be compared with the set of super-symmetric models. 
When we used the
core radius value of $0.21pc$ provided by Gnedin and Ostriker [6] our 
upper limit does not constrain any of the models even when $50\%$ of the 
cluster mass is in the
form of WIMPs.  Recent dynamical studies [10] of M15 and Hubble
Space Telescope observations [8] indicate that the core
radius is much smaller, possibly smaller than 0.03pc. We also calculated
upper limits for such a small core radius and see that the $50\%$ case
starts to conflict with some super-symmetric models.
\section{Conclusion}
Our analysis did not take into account the gamma ray energy.  Using a 
higher threshold should permit to obtain better constraints on heavy WIMPs 
(mass $> 1$ TeV) from the spectral line component. The rates are very 
sensitive to the details of the dark matter distribution in the center of 
the cluster, and our result should become more constraining when we will use  
more realistic radial profile as will be presented at the conference. 
Future ground-based gamma ray detectors 
with lower thresholds and improved discrimination should allow to probe 
annihilating super-symmetric dark matter in globular clusters such as M15 
possibly to the level of few percents of the total cluster mass even for 
low mass WIMPS theories.
\section{Acknowledgments}
We acknowledge the technical assistance of E. Roache and J. Melnick. This 
research is supported by grants from the U.S. Department of Energy, by 
Enterprise Ireland and by PPARC in the UK.
\section{References}
\re
1.\ Baltz, E.A., et al. 2000, Phys. Rev. D 61, 023514

\re
2.\ Bergstr\"om, L. and Gondolo, P. 1996, Astropart.~Phys. 5, 263.

\re
3.\ Berstr\"om, L., Ullio, P. \& Buckley, J.H., 1998, Astropart.~Phys. 9, 137

\re
4.\ Edsj{\"o}, J. and Gondolo, P. 1997, Phys. Rev. D 56, 1879.
\re
5.\ Gerssen, J. et al., 2003, AJ, 125:376-377.
\re
6.\ Gnedin, O.Y., and Ostriker, J.P., 1997, ApJ, 474:223-255.
\re
7.\ P.~Gondolo et al. 2002, astro-ph/0211238.
\re
8.\ Guhathakurta, P. et al., Bulletin of the AAS, Vol. 26, p.1490.
\re 
9.\ LeBohec et al., 2003, in preparation.
\re 
10.\ Murphy, B.W. et al., Bultin of the AAS, Vol. 26, p.1487.
\re
11.\ Moore, B., 1996, ApJ, 461:L13-L16.
\re
12.\ Pryor, C., et al., 1989, AJ, 98, 596.
\re
13.\ Rosenblat, E.I., et al., 1988, ApJ, 330: 191-200.
\re
14.\ Spergel, D.N., et al. 2003, astro-ph/0302209.
\re
15.\ Tyler, C., astro-ph/0203242.
\re
16.\ Urban, M., et al., 1992, Phys. Lett. B, 293, 149.
\re
17.\ Vassiliev, V., et al., 2003 in these proceedings.
\re
18.\ Cohn, H., 1980, ApJ, 242:765-771.
\re
19.\ Gebhardt, K., et al., 1997, AJ, 113, 1026

\endofpaper
\end{document}